\documentclass[twocolumn,showpacs,preprintnumbers,prb,fleqn,floatfix,superscriptaddress]{revtex4}

\usepackage{graphicx}
\usepackage{dcolumn}
\usepackage{bm}

\begin{document}

\title{Influence of the single particle Zeeman energy\\on the quantum Hall ferromagnet at high filling factors}

\author{B. A. Piot and D. K. Maude}
\affiliation{ Grenoble High Magnetic Field Laboratory, Centre
National de la Recherche Scientifique, 25 Avenue des Martyrs,
F-38042 Grenoble, France}
\author{M. Henini}
\affiliation{School of Physics and Astronomy, University of
Nottingham, Nottingham, NG7 2RD, United Kingdom}
\author{Z. R. Wasilewski and J.A. Gupta}
\affiliation{Institute for Microstructural Sciences, National
Research Council, Ottawa, Canada, K1A 0R6}
\author{K. J. Friedland, R. Hey, and K. H. Ploog}
\affiliation{Paul Drude Institut f\"{u}r Festk%
\"{o}rperelektronik, Hausvogteiplatz 5-7, D-10117 Berlin, Germany}
\author{U. Gennser, A. Cavanna, and D. Mailly}
\affiliation{Laboratoire de Photonique et de Nanostructures,
Centre National de la Recherche Scientifique, Route de Nozay,
91460 Marcoussis, France}
\author{R. Airey and G. Hill}
\affiliation{Department of Electronic and Electrical Engineering,
University of Sheffield, Sheffield S1 4DU, United Kingdom}
\date{\today }

\begin{abstract}
In a recent paper [Phys.Rev.B.\textbf{72},245325(2005)], we have
shown that the lifting of the electron spin degeneracy in the
integer quantum Hall effect at high filling factors should be
interpreted as a magnetic-field-induced Stoner transition. In this
work we extend the analysis to investigate the influence of the
single particle Zeeman energy on the quantum Hall ferromagnet at
high filling factors. The single particle Zeeman energy is tuned
through the application of an additional in-plane magnetic field.
Both the evolution of the spin polarization of the system and the
critical magnetic field for spin splitting are well-described as a
function of the tilt angle of the sample in the magnetic field.
\end{abstract}
\pacs{73.43.Qt, 73.43.Nq} \maketitle

The integer quantum Hall effect has historically been described
within the framework of a single electron picture.
Electron-electron interactions are then introduced as a
correction, leading to enhanced spin gaps at odd filling factors.
Clearly, from a perturbation theory point of view, this approach
is wrong, at least for the most widely investigated GaAs system,
for which the energy scale of the electron-electron interactions
($e^2/4\pi \epsilon \ell_{B}$) is more than an order of magnitude
larger than the single particle Zeeman energy ($g^{*}\mu_{B}B$).
At high magnetic field (low filling factors), this has wide
ranging consequences, with the observation of the itinerant
quantum Hall ferromagnet,\cite{Jungwirth} with spin wave
\cite{Bychkov} or spin texture excitations \cite{Fertig} at
filling factor $\nu=1$.

The \emph{collapse} of spin splitting at low magnetic fields (high
filling factors) has been investigated
experimentally\cite{Wong97,Shickler97,Leadley98} and
theoretically.\cite{Fogler95} Leadley and co-workers
\cite{Leadley98} showed that the critical filling factor for the
\emph{collapse} of spin splitting is found to increase with
increasing tilt angle. The Zeeman energy, greatly enhanced at high
tilt angles, favors the transition to a polarized state. This
latter point is theoretically supported by the pioneering work of
Fogler and Shklovskii,\cite{Fogler95} who proposed an order
parameter, $\delta\nu$, to quantitatively characterize the
\emph{collapse} of spin splitting. This order parameter
corresponds to the filling factor difference between two
consecutive resistance maxima in $R_{xx}(B)$, related to spin up
and down sub-levels associated with a given Landau level. In the
Fogler and Shklovskii model, the spin splitting ($\delta \nu$)
collapses, when the disorder broadening of the Landau levels is
comparable to the exchange enhanced spin gap.

In an equivalent, but intuitively different approach, we have
recently shown \cite{PiotPRBsto} that the \emph{appearance} of
spin splitting results from a competition between the disorder
induced energy cost of flipping spins and the exchange energy gain
associated with the polarized state. In this case the Zeeman
energy plays no role, and the only effect of the magnetic field is
to modify the density of states at the Fermi energy, essentially
through the Landau level degeneracy eB/h. Here, we use the
experimental behavior of the order parameter $\delta\nu$ to probe
the appearance of spin splitting in Al$_{x}$Ga$_{1-x}$As/GaAs
heterojunction (HJ) and quantum well (QW) structures. A large
in-plane magnetic field is used to tune (enhance) the single
particle Zeeman energy by more than an order of magnitude.
Experimentally this is achieved by rotating the sample in the
magnetic field. We show that the behavior of the spin polarization
as a function of the tilt angle can be quantitatively described
within the framework of our previously developed approach for the
appearance of spin splitting, with no free
parameters.\cite{PiotPRBsto}

We briefly recall our simple model for the appearance of spin
splitting in the highest occupied Landau level before introducing
the effect of a non-zero Zeeman energy. In the limit of a zero
Zeeman energy, we consider an unpolarized initial state in the
$N^{th}$ Landau level, with a total number of electrons
$n_{tot}=eB/h$. In this situation the Fermi level $E_{F}$ lies in
the center of the degenerate spin up and down sub-levels and the
filling factor of the system is odd. The development of a non-zero
spin polarization requires that the ``disorder'' energy cost of
populating higher energy levels by flipping spin should be less
than the gain in exchange energy stabilizing the newly polarized
state. The energy cost of flipping spin is inversely proportional
to the density of states of one spin sub-level at Fermi level
$D(E_{F})$, and it can be shown\cite{PiotPRBsto} that it will be
energetically favorable for spins to flip when:
 \begin{equation}\label{eqn:eq1}
\\\frac{1}{D(E_{F})}=X,
\end{equation}
where $X$ is the exchange energy between two spins, essentially
depending only on the electron density $n_{s}$. This condition is
nothing other than the well-known Stoner condition for
ferromagnetism in metals.\cite{Stoner}

\begin{figure}[tbp]
\includegraphics[width=0.9\linewidth,angle=0,clip]{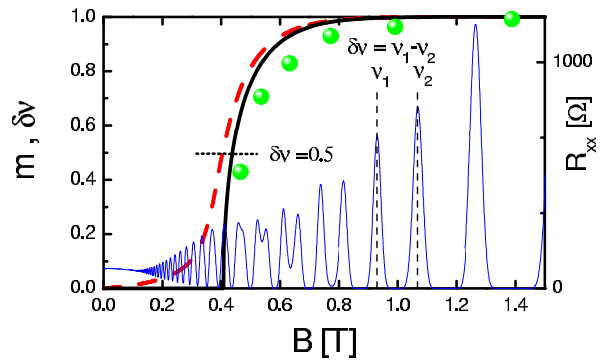}
\caption{(color on-line) $R_{xx}(B)$ measured at $T=50$~mK for
sample NRC0050. The experimentally determined $\delta\nu$ are also
plotted (closed circles). The evolution of the spin polarization
$m(B)$ calculated using Eq.(\ref{eqn:selfconsmGauss}) in the
presence of a Zeeman energy ($g^{*}=-0.44$) (dashed line), and
without Zeeman energy ($g^{*}=0$) (thick solid line) are also
shown. The condition $\delta\nu=0.5$ used to estimate $B_{ss}$ is
indicated by the dotted horizontal line. }\label{fig1}
\end{figure}

In the presence of a non-zero Zeeman energy, there is an initial
spin polarization of the system at odd filling factors. To include
this effect, one has to consider the global spin polarization,
$m$, in Landau level $N$, resulting from the total spin gap
induced by exchange and Zeeman energies. The latter can then be
written,
\begin{equation}\label{eqn:spingap}
\\ \Delta_{s}=|g^{*}|\mu_{B}B+ Xmn_{tot},
\end{equation}
where $g^{*}$ is the effective bare g-factor and $n_{tot}=eB/h$
and $m$ are the occupancy and the spin polarization of the
$N^{th}$ Landau level respectively. Here $m n_{tot}$ corresponds
to the difference between the number of spin up and spin down
electrons which is at the origin of the exchange gap. For
simplicity we define the zero of energy to be at the Fermi level
($E_{F}\equiv 0$). The spin polarization
$m=(n_{\uparrow}-n_{\downarrow})/{n_{tot}}$ resulting from the
spin gap $\Delta_{s}$ can then be written,
\begin{equation}\label{eqn:polarisation}
 m = \frac{1}{n_{tot}}  \int_{-\infty}^{0} \left[D\left(E+ \frac{\Delta_{s}}{2}\right) -D\left(E-
 \frac{\Delta_{s}}{2}\right) \right] dE,
\end{equation}
where $D(x)=(1/\Gamma\sqrt{\pi})\exp (-x^{2}/\Gamma^{2})$ for
Gaussian broadened Landau levels of width $\Gamma$.
Eqs.(\ref{eqn:spingap}) and (\ref{eqn:polarisation}) need to be
solved self-consistently to find $m$ and $\Delta_{s}$ and are
essentially equivalent to the solution proposed by Fogler and
Shklovskii.\cite{Fogler95} For Gaussian Landau levels,
\begin{equation}\label{eqn:selfconsmGauss}
\\ m =erf\left(\frac{1}{2}\frac{(|g^{*}| \mu_{B} B+ X m n_{tot})}{\Gamma}\right),
\end{equation}
where $erf(x)=\frac{2}{\sqrt{\pi}}\int_{0}^{x}e^{-t^{2}}dt $. We
note that assuming an energy independent density of states around
$E_{F}$ (equivalent to assuming a ``rectangular'' Landau level) we
obtain,
\begin{equation}\label{eqn:squareapprox}
\\ m=\frac{(|g^{*}| \mu_{B} B + X m n_{tot}) D(E_{F}) }{n_{tot}}.
\end{equation}
However, this is only a good approximation for small values of
$m$, so the equation required to express the continuous evolution
of the spin polarization as a function of the magnetic field is
Eq.(\ref{eqn:selfconsmGauss}). Fogler and Shklovskii
\cite{Fogler95} have shown that there is a simple relation,
$m=\delta \nu$, linking the polarization and the filling factor
separation of the peaks in $R_{xx}(B)$, which allows a direct
comparison of theory with experiment. In Fig.~(\ref{fig1}) we plot
the magnetoresistance $R_{xx}(B)$ and measured $\delta\nu$ for
sample NRC0050 (the sample parameters are summarized in
Table~\ref{tab:table1}). The evolution of $m(B)$ calculated using
Eq.(\ref{eqn:selfconsmGauss}) with $g^{*}=-0.44$, the generally
accepted value for bulk GaAs, \cite{Weisbuch} and $g^{*}=0$, are
also plotted in Fig.\ref{fig1}. The values used for $\Gamma$ and
$X$ used in the calculation have been independently determined for
this sample as explained in Ref.~[\onlinecite{PiotPRBsto}].

\begin{table}[b]
\centering \caption{Parameters of the samples
investigated.}\label{tab:table1}
\begin{tabular}{c|cccc}
Sample & $n_{s}$~(cm$^{-2}$) & $\Gamma$~(K)& Structure & $g^{*}$\\
\hline
 NRC0050 & $1.7 \times 10^{11}$ & $1.8 \pm 0.1$ & HJ & -0.44\\
 LPN06 & $4.0 \times 10^{11}$ & $3.9 \pm 0.6$ & HJ & -0.44\\
 NU1783b & $1.8 \times 10^{11}$ & $1.8 \pm 0.2$ & HJ & -0.44\\
 F1200 & $7.6 \times 10^{11}$ & $1.7 \pm 0.1$ & QW & -0.1\\
\end{tabular}
\end{table}

Eq.(\ref{eqn:selfconsmGauss}) reproduces extremely well the
collapse of the spin polarization observed in $\delta\nu$,
especially considering that there are no fitting parameters. The
effect of a non-zero Zeeman energy is similar to the one obtained
in Ref.~[\onlinecite{Fogler95}], shifting the phase transition to
lower magnetic field. To evaluate quantitatively the Zeeman
correction, we define a critical magnetic field, $B_{ss}$,
corresponding to a value of $\delta\nu=0.5$, as already proposed
in Refs.~[\onlinecite{Leadley98}] and [\onlinecite{Fogler95}].
$B_{ss}$ can be extracted from the model by putting $m= \delta
\nu=0.5$ into Eq.(\ref{eqn:selfconsmGauss}), with and without the
Zeeman correction. The difference between these two results is at
most $\sim 10$\% for $g^{*}=-0.44$, undiscernible within
experimental error, confirming the negligible role of the Zeeman
energy in the perpendicular configuration.

It is however possible to increase the Zeeman energy by applying a
strong in-plane magnetic field. Experimentally this can be
achieved by rotating the sample away from the $B$ normal to the 2D
electron gas configuration. The Zeeman energy depends on the total
magnetic field $B$, in contrast to orbital effects such as the
Landau level degeneracy or the cyclotron energy which are only
sensitive to the $B_{\perp}$ component of the field.  Hence
rotation can be used to tune the Zeeman energy. For a given
perpendicular magnetic field $B_{\perp}$, the strength of the
Zeeman energy can be increased by orders of magnitude at high tilt
angles. We therefore expect that rotating the sample should
provide an incisive test of the zero free parameter model
developed to predict the appearance of spin splitting in
Ref.~[\onlinecite{PiotPRBsto}].

\begin{figure}[tbp]
\includegraphics[width=0.8\linewidth,angle=0,clip]{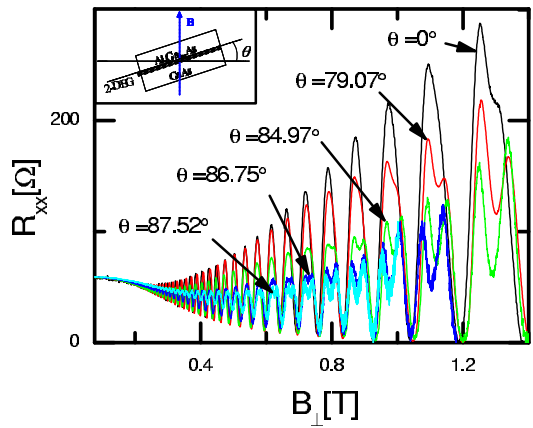}
\caption{(color on-line) $R_{xx}(B_{\perp})$ for sample $LPN06$
measured at $T=30$~mK for different tilt angles
($\theta$).}\label{fig2}
\end{figure}

\begin{figure}[tbp]
\includegraphics[width=0.8\linewidth,angle=0,clip]{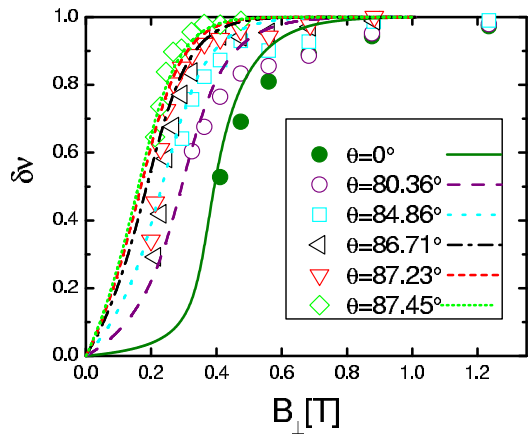}
\caption{(color on-line) Parameter $\delta\nu (B_{\perp})$ for
sample $NU1783b$, extracted from measured $R_{xx}(B)$ at $T=50mK$,
for different tilt angles (symbols). Spin polarization $m$
calculated using Eq.(\ref{eqn:selfconsmGaussteta}) (solid and
broken lines).}\label{fig3}
\end{figure}

The effect of the increased single particle Zeeman gap is clearly
visible in Fig. \ref{fig2}, in which plot $R_{xx}(B_{\perp})$
measured at $T=30$~mK for a GaAs heterojunction at various tilt
angles ($\theta$). $R_{xx}(B_{\perp})$ was measured using a
standard low frequency lock-in technique under magnetic fields up
to $23T$ in a dilution fridge equipped with an \emph{in-situ}
rotating sample holder. As the tilt angle is increased, for a
given $B_{\perp}$, the minima at odd filling factor strengthen,
reflecting the increase of the spin gap. This effect can also be
seen in the behavior of $\delta\nu$ as a function of $B_{\perp}$
(extracted from $R_{xx}(B_{\perp})$),  for different tilt angles,
plotted for a GaAs heterojunction in Fig.\ref{fig3}. The
appearance of spin splitting is clearly shifted to lower
$B_{\perp}$ with increasing Zeeman energy at high tilt angles.
Tilting the sample increases the Zeeman energy without affecting
either the disorder or the exchange parameter. This is quite clear
from Eq.(\ref{eqn:selfconsmGauss}) where the total magnetic field
only enters through the Zeeman term, with $n_{tot}$ involving only
the perpendicular component. A similar approach was first proposed
in Ref.~[\onlinecite{Leadley98}] to extract the enhanced g-factor
from the coincidence method at high tilt angles. Practically,
Eq.(\ref{eqn:selfconsmGauss}) can in this case be written,

\begin{equation}\label{eqn:selfconsmGaussteta}
\\ m =erf\left(\frac{1}{2\Gamma}\left(|g^{*}|\mu_{B}\frac{B_{\perp}}{\cos(\theta)}+
Xm\frac{eB_{\perp}}{h}\right)\right).
\end{equation}

The predicted behavior is shown in Fig.\ref{fig3}, calculated
using Eq.(\ref{eqn:selfconsmGaussteta}), with the parameters
$\Gamma$ and $X(n_{s})$, determined from an analysis of the
oscillations in $R_{xx}(B)$ before spin splitting occurs, and, the
calculations of Attacalite et al., \cite{Attaccalite}
respectively, as detailed in Ref.~[\onlinecite{PiotPRBsto}]. We
impose here the generally accepted value of $g^{*}=-0.44$ for bulk
GaAs. The effect of an increasing tilt angle on the collapse of
$\delta\nu$ is well reproduced, considering the slight discrepancy
between our model and experiment in the perpendicular
configuration (see the curve for $\theta=0^{\circ}$ and the data
associated (full circles)). We stress that there are no free
fitting parameters in our model which nevertheless provides a good
quantitative description. For high tilt angles, the experimental
collapse of $\delta\nu$ seems more pronounced than the predicted
variation and a possible reason for this is proposed later.

It is important to mention here that the self consistent nature of
Eq. (\ref{eqn:selfconsmGaussteta}), which arises from the
dependence of the exchange gap on the spin polarization, is
essential in obtaining such good agreement. The reason for this is
that even at fixed perpendicular magnetic field, the spin
polarization at odd filling factor can be modified due to the
change in the single particle Zeeman energy, if the spin Landau
levels overlap, leading to a modification of the exchange gap.

\begin{figure}[tbp]
\includegraphics[width=0.8\linewidth,angle=0,clip]{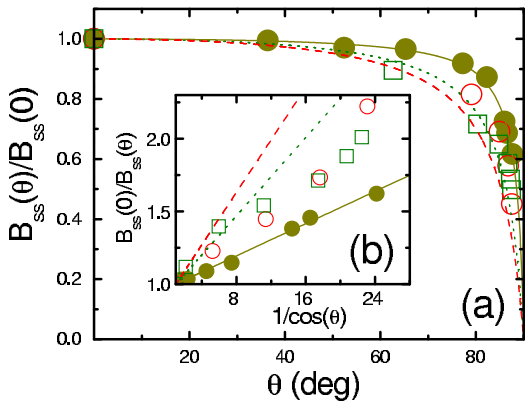}
\caption{(color on-line) (a) $B_{ss}(\theta)/B_{ss}(0)$ as a
function of $\theta$ for 3 different samples at $T=50$~mK: $F1200$
(full circles), $NU1783b$ (open squares) and $LPN06$ (open
circles). Self-consistent solutions of
Eq.(\ref{eqn:selfconsmGaussteta}), for the three samples, are also
plotted (lines). (b) $B_{ss}(0)/B_{ss}(\theta)$ as a function of
$1/cos(\theta)$ for the three samples (symbols) together with the
predictions of Eq.(\ref{eqn:selfconsmGaussteta})
(lines).}\label{fig4}
\end{figure}

We now turn to the evolution of the critical magnetic field
$B_{ss}$ with tilt angle. From the curves of Fig.\ref{fig3} we can
extract the experimental magnetic field $B_{ss}$ corresponding to
the condition $\delta\nu=0.5$. This quantity is shown for the 3
samples for which we have rotation data in Fig.\ref{fig4} as a
function of the tilt angle $\theta$. $LPN06$ and $NU1783b$ are the
two GaAs heterojunction already presented and $F1200$ is a 13nm
wide GaAs quantum well. In order to focus only on the effect of
tilting, we plot the value $B_{ss}(\theta)$ normalized by its
value in the perpendicular configuration, $B_{ss}(0)$. As
expected, $B_{ss}(\theta)/B_{ss}(0)$ is greatly reduced at high
tilt angles, when the Zeeman energy is increased by over an order
of magnitude. Reductions of more than $50\%$ are observed  for
angles approaching $90^{\circ}$. From a self-consistent solution
of Eq.(\ref{eqn:selfconsmGaussteta}) we can obtain the predicted
magnetic field $B_{ss}$ as a function of the angle $\theta$. As
$B_{ss}$ corresponds to the condition $\delta\nu=0.5$, the
solution is obtained setting $m=0.5$ in
Eq.(\ref{eqn:selfconsmGaussteta}). The predicted evolution of
$B_{ss}(\theta)/B_{ss}(0)$ is plotted, as solid and broken lines,
for the three samples in Fig.\ref{fig4}(a).

As before, $\Gamma$ and $X$ have been independently determined for
each sample.\cite{PiotPRBsto} For the heterojunction samples,
$NU1783b$ and $LPN06$, good agreements with the data are obtained
using the generally accepted value of $g^{*}=-0.44$ for bulk GaAs.
\cite{Weisbuch} For the 13nm quantum well sample, $F1200$, we
expect a significantly lower value for the bare g-factor owing to
the confinement energy and the penetration of the wave function
into the barriers.\cite{Ivchenko} In addition, non-parabolicity is
expected to further reduce the g-factor\cite{Dobers88} as the
electronic density (Fermi energy) is in this sample quite high
($n_{s}=7.5\times10^{11}cm^{-2}$). A good agreement with the data
can be obtained using $g^{*}=-0.1$ which is not unreasonable.

As can be seen in Fig.\ref{fig4}(a) a generally good quantitative
agreement is obtained confirming the relevance of the physical
approach proposed. A slight discrepancy ($\sim 20-30$\%) is
observed at large angles for samples $NU1783b$ and $LPN06$ between
the prediction and the experimental $B_{ss}$, the latter being
slightly larger. This can be seen in Fig.\ref{fig4}(b) in which we
plot the experimental and predicted $B_{ss}(0)/B_{ss}(\theta)$ as
a function of $1/cos(\theta)$. Intuitively, the linear dependence,
observed for the exact numerical solution of
Eq.(\ref{eqn:selfconsmGaussteta}), can be understood from the
approximate expression for $m$ in Eq.(\ref{eqn:squareapprox}),
writing $B=B_{\perp}/\cos(\theta)$, $D(E_{F})=eB_{\perp}/h\Gamma
\sqrt(\pi)$, $n_{tot}=eB_{\perp}/h$, and $m=0.5$ which gives,

\[\frac{B_{ss}(0)}{B_{ss}(\theta)} \approx \frac{1}{\cos
(\theta)}\left(\frac{g^{*}\mu_{B}}{g^{*}\mu_{B} + Xe/2h}\right) +
\left(\frac{Xe/2h}{g^{*}\mu_{B} + Xe/2h}\right)\]

This expression provides a reasonable prediction for the slope of
the $1/\cos(\theta)$ dependence, for decreases in $B_{ss}(\theta)$
of less than $\sim 50$\%.  A deviation from theory is clearly
visible at large $\theta$ (large $1/\cos(\theta$)) for samples
$NU1783b$ and $LPN06$. For $F1200$ however a linear behavior in
good agreement with the prediction is observed. A possible
explanation for these discrepancies is that the large in-plane
magnetic field, at large tilt angles, increases the Landau level
width, which would shift the appearance of spin splitting to
higher perpendicular magnetic field. At large tilt angles, the
amplitude of the oscillations in $R_{xx}(B_{\perp})$ decrease
significantly for a given $B_{\perp}$ (see Fig.\ref{fig2}),
revealing a modification of the scattering as the in-plane
magnetic field increases. The effect of the in-plane magnetic
field, which we have also observed in in-plane magnetoresistance
measurements ($\theta=90^{\circ}$), is in GaAs a complex interplay
between spin and orbital effects \cite{DasSarma,Tutuc} affecting
both the quantum lifetime (Landau level width) and the effective
mass. The orbital part of this effect is weaker in quantum wells
in which the spacing between electronic subbands is larger than in
heterojunctions, limiting the inter-subband transitions induced by
an in-plane magnetic field. This would be consistent with the fact
that sample $F1200$, a $13nm$ quantum well, is not affected by
this process and shows no deviation from the predicted linear
behavior.

In summary, we have extended our model for the quantum Hall
ferromagnet at high filling factors to the case of a non-zero
Zeeman energy. The Zeeman energy has been tuned via tilted field
measurements. Our simple model, with no free fitting parameters,
provides a reasonable quantitative description of the Zeeman
energy dependance of spin polarization (spin splitting) at odd
filling factors.


\end{document}